\documentclass[aps,prx,fixfloat,twocolumn]{revtex4-2}
\usepackage{graphicx,amsmath,amsfonts,bm, color}
\usepackage{wasysym}
\usepackage{xcolor,hyperref}
\hypersetup{
   colorlinks,
   linkcolor={blue!50!black},%{red!80!black},
   citecolor={blue!50!black},
   urlcolor={blue!80!black}
}
\usepackage{enumitem}

\renewcommand{\=}{\!=\!}

\DeclareMathAlphabet{\mathitbf}{OML}{cmm}{b}{it}

\newcommand{\calBold}[1]{\mbox{\boldmath${\cal #1}$}}

\begin{document}

\title{The boson peak in the vibrational spectra of glasses}
\author{Avraham Moriel$^{1}$}
\author{Edan Lerner$^{2}$}
\author{Eran Bouchbinder$^{1}$}
\email{eran.bouchbinder@weizmann.ac.il}
\affiliation{$^{1}$Chemical and Biological Physics Department, Weizmann Institute of Science, Rehovot 7610001, Israel\\
$^{2}$Institute for Theoretical Physics, University of Amsterdam, Science Park 904, Amsterdam, Netherlands}

\begin{abstract}
A hallmark of glasses is an excess of low-frequency, nonphononic vibrations, in addition to phonons. It is associated with the intrinsically nonequilibrium and disordered nature of glasses, and is generically manifested as a THz peak --- the boson peak --- in the ratio of the vibrational density of state (VDoS) and Debye's VDoS of phonons. Yet, the excess vibrations and the boson peak are not fully understood. Here, using reanalysis of experimental data, extensive computer simulations and a mean-field model, we show that the nonphononic part of the VDoS itself features both a universal power-law tail and a peak, entirely accounted for by quasi-localized nonphononic vibrations, whose existence was recently established. We explain the mild variation of the peak's frequency and magnitude with glasses' thermal history, along with the strong variation of the power-law tail. We also show that modes that populate the peak's region feature many coupled quasi-localized nonphononic vibrations, when their spatial structure is considered. Our results provide a unified physical picture of the low-frequency vibrational spectra of glasses, and in particular elucidate the origin, nature and properties of the boson peak.
\end{abstract}

\maketitle

\section{Introduction}

The glassy state of matter, generically formed by quickly cooling a liquid to avoid crystallization, still poses fundamental scientific challenges~\cite{falk_review_2016,Ediger_review_2017,elasto_plastic_review_2018,parisi2020theory,ramos2023low}. The fast cooling leads to a self-organized disordered solid that lacks the long-range order of crystalline solids~\cite{Cavagna200951} --- a glass. The disordered and nonequilibrium nature of glasses endows them with unique physical properties, different from their crystalline counterparts. Of particular importance are the low-frequency vibrational spectra, which control various mechanical, transport and thermodynamic properties of solids~\cite{Zeller_and_Pohl_prb_1971,phillips1972tunneling,anderson1972anomalous,buchenau_prb_1992,pohl_review,david_fracture_mrs_2021,gzregorz_attenuation_jcp_2022}.

The low-frequency vibrational spectra of solids --- either crystalline or glassy --- contain phonons, which are extended vibrations emerging due to global symmetries, independently of the underlying material structure~\cite{kittel2005introduction}. Low-frequency phonons are well described by Debye’s vibrational density of state (VDoS), ${\cal D}_{\rm D}(\omega)\=A_{\rm D}\,\omega^2$ (in three dimensions), where $\omega$ is the vibrational (angular) frequency and $A_{\rm D}$ is a prefactor that depends on the elastic properties of the solid. The low-frequency vibrational spectra of glasses, however, are known to universally feature also other, nonphononic, modes. This is commonly (but not exclusively~\cite{ruocco_boson_peak_ruocco_2006,KALAMPOUNIAS20064619}) established by dividing the VDoS (measured by various scattering techniques~\cite{HUDSON200625,weber2000raman}), ${\cal D}(\omega)$, by Debye's phononic VDoS, ${\cal D}_{\rm D}(\omega)\!\sim\!\omega^2$. The reduced VDoS ${\cal D}(\omega)/\omega^2$ universally deviates from a constant at low frequencies and features a peak in the THz regime, known as the boson peak --- a hallmark of glasses~\cite{sokolov_1986,soft_potential_model_1991,ramos1997quantitative,wischnewski1998neutron,kojima1999boson,surovtsev2000inelastic,surovtsev2003density,duval2003effect,parisi_boson_peak_2003,wyart_epl_2005,monaco2006density,monaco2006effect,baldi2023vibrational,Schirmacher_2013_boson_peak,eric_boson_peak_emt}. Yet, despite decades of research, the origin, nature and properties of the nonphononic boson peak vibrations remain highly debated~\cite{Schober_Oligschleger_numerics_PRB,Gurevich2003,Gurevich2007,Monaco_prl_2011,Lubchenko1515,tanaka_2d_modes_2022,boson_peak_2d_arXiv2022,gonzalez2023understanding}.

Significant recent progress~\cite{modes_prl_2016,ikeda_pnas,modes_prl_2018,LB_modes_2019,modes_prl_2020,JCP_Perspective} elucidated the nature and properties of the low-frequency tail, $\omega\!\to\!0$, of the nonphononic part of the VDoS ${\cal D}_{\rm G}(\omega)$. By disentangling extended phononic and quasi-localized nonphononic vibrations, the latter were shown to follow a universal non-Debye VDoS ${\cal D}_{\rm G}(\omega)\=A_{\rm g}\,\omega^4$ in the $\omega\!\to\!0$ limit, below the boson peak~\cite{modes_prl_2016,JCP_Perspective}. Here, $A_{\rm g}$ is a nonuniversal prefactor that depends on the glass nonequilibrium history and its emerging disordered state~\cite{pinching_pnas,JCP_Perspective}. These quasi-localized nonphononic vibrations feature a localized core of linear size of about 10 atoms, where displacement amplitudes are large and highly disordered in orientation, and power-law decaying displacements away from the core~\cite{JCP_Perspective}, see Fig.~\ref{fig:fig1}a.

Here, using experimental data, extensive atomistic computer simulations and solutions of a mean-field model of interacting glassy vibrations, we show that the nonphononic part of the VDoS ${\cal D}_{\rm G}(\omega)$ features both a universal $\sim\!\omega^4$ tail and an intrinsic peak, entirely accounted for by quasi-localized nonphononic vibrations. The peak's frequency and magnitude mildly increase with decreasing state of glassy disorder (e.g.~controlled by thermal annealing), while the $\sim\!\omega^4$ tail is strongly suppressed, in agreement with the predictions of the mean-field model of interacting quasi-localized vibrations. The model thus provides a unified picture of the low-frequency properties of the nonphononic VDoS ${\cal D}_{\rm G}(\omega)$, both the universal $\sim\!\omega^4$ tail and the peak regime. Finally, we predict the number of coupled quasi-localized nonphononic vibrations spatially composing the peak's modes as a function of $\omega$, in line with very recent observations. Overall, our results elucidate the origin, nature and salient properties of the boson peak in glassy solids.

\section{The form of the nonphononic part of the VDoS}

\begin{figure*}[ht!]
\center
\includegraphics[width=1\textwidth]{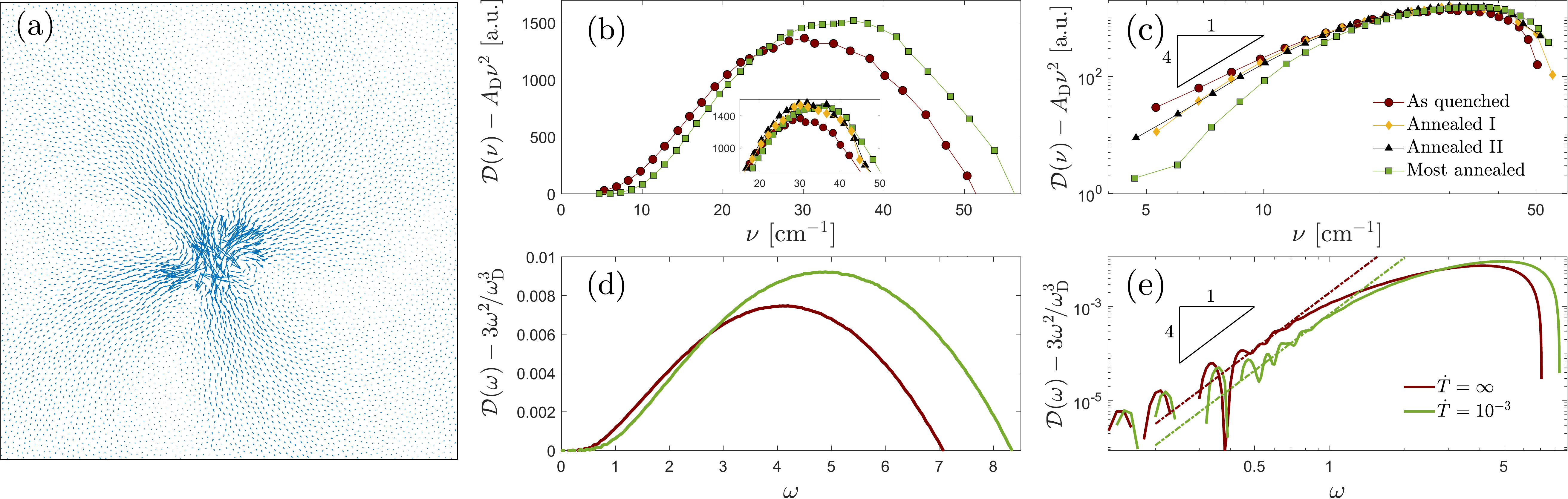}
\caption{(a) A quasi-localized vibration extracted from the universal $\sim\!\omega^4$ tail of the nonphononic VDoS of a two-dimensional computer glass~\cite{JCP_Perspective}. The arrows represent the atomic displacements within the vibrational mode, which features a localized core of linear size of about 10 atoms, where displacement amplitudes are large and highly disordered in orientation. Away from the core, displacements decay as a power-law and feature a quadrupolar azimuthal dependence. (b) The nonphononic VDoS ${\cal D}_{\rm G}(\nu)\!\equiv\!{\cal D}(\nu)\!-\!A_{\rm D}\,\nu^2$ of boron-oxide (B$_2$O$_3$) glass samples, reported in~\cite{surovtsev2003density} for different thermal annealing conditions. Shown are the data for the as quenched and most annealed samples (see legend on panel (c)). In~\cite{surovtsev2003density}, ${\cal D}(\nu)/\nu^2$ (note ${\cal D}(\nu)$ is denoted as $g(\nu)$ therein) and $A_{\rm D}$ are separately reported, but not ${\cal D}_{\rm G}(\nu)$ itself. ${\cal D}_{\rm G}(\nu)$ reveals a peak, which increases in magnitude and shifts to higher frequencies with increased thermal annealing (corresponding to less disordered glassy states). Data for intermediate annealed samples are added in the inset, see legend in panel (c). (c) The same as the inset of panel (b), but on a double-logarithmic scale. The small $\nu$ tail approximately reveals the universal $\sim\!\nu^4$ behavior (see power-law triangle), in the most pronounced manner for the two intermediate annealed sample. Moreover, the amplitude of the tail is strongly suppressed upon annealing, in sharp contrast to the mild increase $\nu_{\rm p}$ and ${\cal D}_{\rm G}(\nu_{\rm p})$ observed in panel (b). (d) ${\cal D}_{\rm G}(\omega)\!=\!{\cal D}(\omega)-3\omega^2/\omega^3_{\rm D}$ using a canonical computer glass made of 4 million atoms~[supplementary material
(SM) section S-1B], where $\omega_{\rm D}$ is Debye's frequency and $A_{\rm D}\!=\!3/\omega^3_{\rm D}$. The two curves correspond to different thermal histories, the upper one to an instantaneous quench through the glass transition, $\dot{T}\!=\!\infty$, and the lower to a smaller quench rate, $\dot{T}\!=\!10^{-3}$, see legend on panel (e). All quantities are reported in simulational units (SM S-1B). (e) The same as panel (d), but on a double-logarithmic scale. The dashed lines of slope 4 (see power-law triangle) are guides to the eye that indicate the universal $\sim\!\omega^4$ tail. At the lower end of the VDoS, relics of discrete phonon bands are observed, which are a finite-size effect~\cite{phonon_widths}. Comparing the simulational results in panels (d)-(e) to the experimental results in panels (b)-(c) reveals striking similarities, despite the inherent size limitation in the former.}
\label{fig:fig1}
\end{figure*}

The common practice is to characterize the boson peak through the reduced VDoS ${\cal D}(\omega)/\omega^2$, where ${\cal D}_{\rm D}(\omega)\!\sim\!\omega^2$ stands for Debye's VDoS of phonons. However, in view of the major recent progress in understanding the universal $\sim\!\omega^4$ tail of the nonphononic VDoS ${\cal D}_{\rm G}(\omega)$, populated by quasi-localized vibrations (see Fig.~\ref{fig:fig1}a), we follow~\cite{ruocco_boson_peak_ruocco_2006,KALAMPOUNIAS20064619} and shift our focus to ${\cal D}_{\rm G}(\omega)$, aiming to understand its generic properties also above the $\sim\!\omega^4$ tail. ${\cal D}_{\rm G}(\omega)$ can be extracted according to ${\cal D}_{\rm G}(\omega)\={\cal D}(\omega)\!-\!A_{\rm D}\,\omega^2$, implying that as far as the number of vibrational modes per frequency $\omega$ is concerned, quasi-localized nonphononic vibrations and phonons make additive contributions to ${\cal D}(\omega)$ (while it is well established that phononic and nonphononic modes do hybridize/mix in space~\cite{phonon_widths,boson_peak_2d_arXiv2022}). The importance of considering ${\cal D}_{\rm G}(\omega)$ (termed e-VDoS in~\cite{ruocco_boson_peak_ruocco_2006}) instead of ${\cal D}(\omega)/\omega^2$ has been highlighted in~\cite{ruocco_boson_peak_ruocco_2006}, along with the associated experimental difficulties. In particular, the challenge is to obtain experimental measurements of ${\cal D}(\omega)$ in glassy samples where disorder is systematically controlled, and where the prefeactor $A_{\rm D}$ in Debye's VDoS of phonons ${\cal D}_{\rm D}(\omega)\=A_{\rm D}\,\omega^2$ is independently extracted.
%%%%%%%%%%%%%%%%%%%%%%%%%%%%%%%%%%%%%%%
%%%%%%%%%%%%%%%%%%%%%%%%%%%%%%%%%%%%%%%

The experiments of~\cite{surovtsev2003density} report on the VDoS of boron-oxide (B$_2$O$_3$) glass samples for different thermal histories, as well as on Debye's prefactor $A_{\rm D}$, and hence are particularly important here. These experiments use low-frequency Raman scattering to measure the VDoS~\cite{weber2000raman}. Vibrational spectra based on Raman scattering are commonly reported in terms of the Raman wavenummber shift $\nu$ (in units of cm$^{-1}$), rather than in terms of $\omega$, though the two are linearly related (in particular, $\omega\=1$ THz corresponds to $\nu\=33.3$ cm$^{-1}$). Consequently, the vibrational spectra are commonly reported as ${\cal D}(\nu)$ (note that~\cite{surovtsev2003density} employ the notation $g(\nu)$, not used here).

Different thermal histories are realized by subjecting as quenched B$_2$O$_3$ glass samples to various annealing treatments, i.e.~annealing at different temperatures in the vicinity of the glass temperature $T_{\rm g}$ for different times. The annealed samples give rise to less disordered, denser glassy states compared to the as quenched sample. Importantly, in addition to measuring ${\cal D}(\nu)$ for each glass sample, Debye's prefactor $A_{\rm D}$ was also extracted (SM S-1A). In Fig.~\ref{fig:fig1}b, we present ${\cal D}_{\rm G}(\nu)\!\equiv\!{\cal D}(\nu)\!-\!A_{\rm D}\,\nu^2$ for the as quenched and most annealed samples (see legend on panel (c)). First, we observe that ${\cal D}_{\rm G}(\nu)$ features a rather symmetric peak, without invoking any phonons, in agreement with~\cite{surovtsev2003density}. Second, the peak's frequency $\nu_{\rm p}$ and magnitude ${\cal D}_{\rm G}(\nu_{\rm p})$ mildly increase with annealing (by $20\%$ and $11\%$ relative to the as quenched sample, respectively, see SM S-1A). The inset also presents data for two intermediate annealed samples, see legend in panel~(c).

In Fig.~\ref{fig:fig1}c, we present the experimental data shown in the inset of Fig.~\ref{fig:fig1}b on a double-logarithmic scale. Remarkably, the small $\nu$ tail approximately reveals the universal $\sim\!\nu^4$ behavior, in the most pronounced manner for the two intermediate curves. To the best of our knowledge, this is the first direct experimental evidence for the universal $\sim\!\omega^4$ tail of the nonphononic VDoS of glasses. Moreover, the tail is strongly suppressed upon annealing. In particular, ${\cal D}_{\rm G}(\nu)$ of the most annealed sample is an order of magnitude smaller than that of the as quenched sample at the lowest $\nu$ available, in sharp contrast to the mild increase in $\nu_{\rm p}$ and ${\cal D}_{\rm G}(\nu_{\rm p})$.

We next set out to test whether these remarkable experimental observations are reproduced in atomistic computer glasses. To that aim, we considered a canonical computer glass-forming model in three dimensions (SM S-1B). We use systems of 4 millions atoms, being large enough to restrict the emergence of discrete phonon bands in the VDoS ${\cal D}(\omega)$ --- which are a finite-size effect~\cite{phonon_widths} (whose relics are seen in Fig.~\ref{fig:fig1}e) entirely absent from experimental spectra --- yet small enough to be computationally feasible and to induce some variability in their degree of structural disorder (SM S-1B and S-2A). The latter is achieved by considering an instantaneous quench, leading to more disordered glassy states, and a finite quench, leading to more ordered glassy states (see figure legend). In Figs.~\ref{fig:fig1}d-e, we present ${\cal D}(\omega)\!-\!3\omega^2/\omega^3_{\rm D}$ (where $A_{\rm D}\=3/\omega^3_{\rm D}$, with $\omega_{\rm D}$ being Debye's frequency) in double-linear (panel d) and double-logarithmic (panel e) scales. The salient features of the experimental observations in Figs.~\ref{fig:fig1}b-c are reproduced by the computer simulations, despite the inherent size limitation (e.g.~setting a bound on the lowest cooling rate), see SM S-2A.

Our next goal is to theoretically understand the main experimental (and simulational) observations. These include the existence of a peak in ${\cal D}_{\rm G}(\omega)$ above the $\sim\!\omega^4$ tail, the mild increase in the peak's frequency $\omega_{\rm p}$ and its magnitude ${\cal D}_{\rm G}(\omega_{\rm p})$ with thermal annealing, along with the corresponding strong suppression of the $\sim\!\omega^4$ tail. We achieve this by studying a mean-field model of quasi-localized nonphononic vibrations.

\section{A mean-field model of quasi-localized nonphononic vibrations}

In order to develop a theoretical understanding of the above discussed observations, we build on a recently formulated mean-field model of quasi-localized nonphononic vibrations~\cite{scipost_mean_field_qles_2021,meanfield_qle_pierfrancesco_prb_2021} that reproduced the ${\cal D}_{\rm G}(\omega)\=A_{\rm g}\omega^4$ VDoS in the $\omega\!\to\!0$ tail below the boson peak, as well as the dependence of the prefactor $A_{\rm g}$ on the disorder parameters of the model. The model envisions small groups of atoms/molecules in an instantaneous snapshot of the liquid state, prior to cooling/quenching through the glass temperature, which feature a collective vibration characterized by a stiffness (spring constant) $\kappa_i$, where $i$ is an index of the group (for $\kappa_i\!>\!0$, the vibrational frequency is $\omega_i\=\sqrt{\kappa_i}$). Since an instantaneous liquid state features both negative (unstable) and positive local stiffnesses, the probability to find a vanishing $\kappa_i$ is finite. Consequently, at small positive stiffnesses (frequencies), e.g.~within $[0,\, \kappa_0]$, the probability to observe a stiffness $\kappa$ is given by $p(\kappa)\=1/\kappa_0$, to leading order. The latter implies a liquid-like VDoS $g_0(\omega)\=2\omega/\kappa_0\!\equiv\!2\omega/\omega_0^2$.
%%%%%%%%%%%%%%%%%%%%%%%%%%%%%%%%%%%%%%%%%%%%%%%%%%%%
\begin{figure}[ht!]
\center
\includegraphics[width=0.48\textwidth]{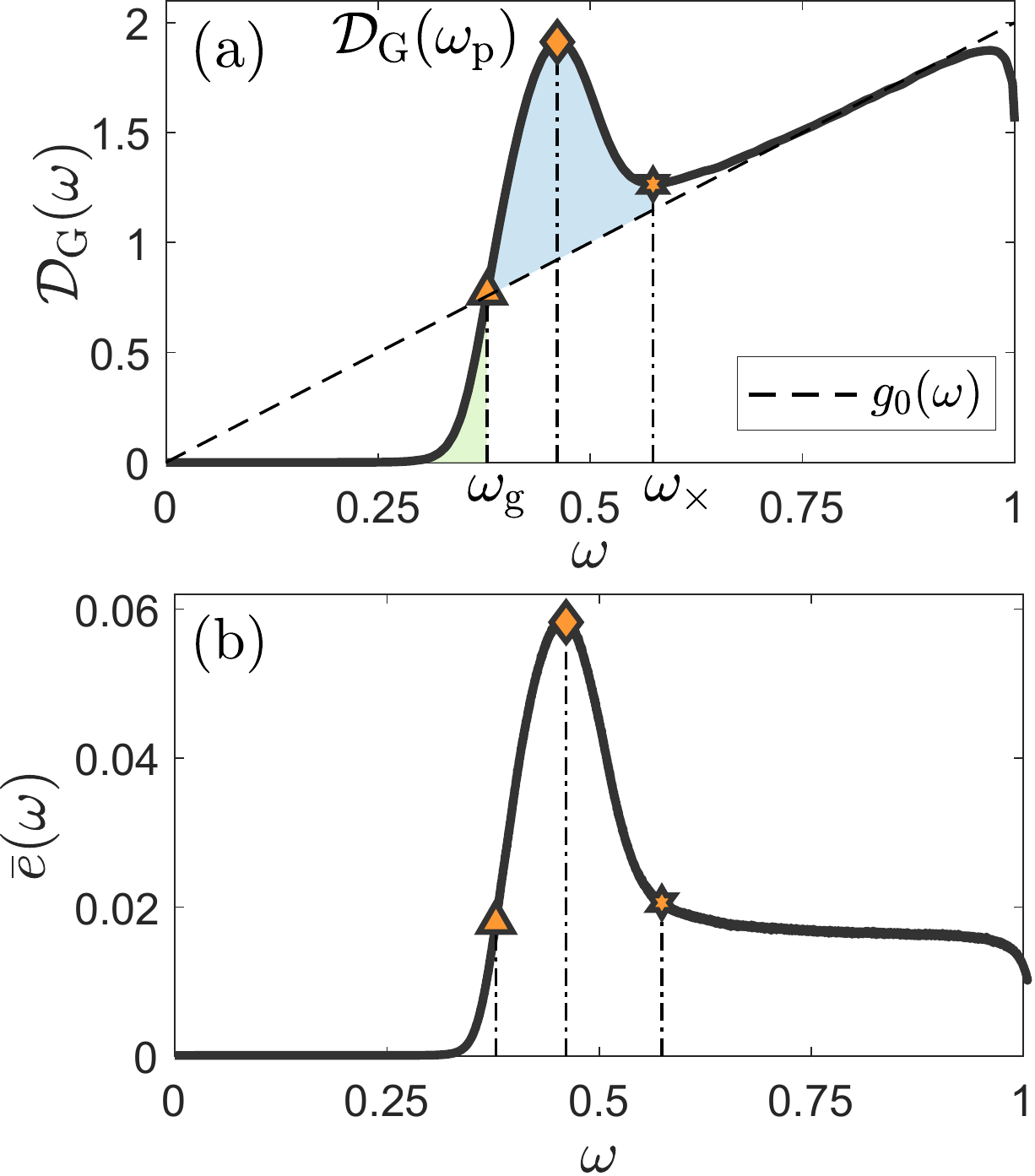}% Here is how to import EPS art
\caption{(a) A representative nonphononic VDoS ${\cal D}_{\rm G}(\omega)$ of the mean-field model of Eq.~\eqref{eq:Hamiltonian}, in the small $y\!=\!J/(h^{1/3}\kappa_0^{1/2})$ regime (thick solid line), see (SM S-1C for details). Here and elsewhere in this work, we set $\kappa_0\!=\!1$ (which implies that the largest frequency is $\omega_0\!=\!\sqrt{\kappa_0}\!=\!1$, as shown). ${\cal D}_{\rm G}(\omega)$ attains a local maximum at $\omega_{\rm p}$, as marked (orange diamond). The thin dashed line corresponds to the initial liquid-like VDoS $g_0(\omega)\!=\!2\omega/\omega_0^2$ (see legend). $\omega_\times$ (orange star, corresponding to the local minimum of ${\cal D}_{\rm G}(\omega)$) marks the frequency above which $g_0(\omega)$ and ${\cal D}_{\rm G}(\omega)$ approximately coincide. $\omega_{\rm g}$ (orange triangle) is a lower intersection frequency of the two curves. The two (blue and green) light shaded regions are discussed in the text. (b) The corresponding averaged participation ratio $\bar{e}(\omega)$, see text for discussion.
\label{fig:fig2}}
\end{figure}
%%%%%%%%%%%%%%%%%%%%%%%%%%%%%%%%%%%%%%%%%%%%%%%%%%%%

Describing every liquid-like vibration by a single collective coordinate $x_i$, and considering the lowest stabilizing anharmonicity~\cite{scipost_mean_field_qles_2021,meanfield_qle_pierfrancesco_prb_2021}, each vibration $i$ is effectively an anharmonic oscillator with energy $\kappa_i x_i^2/2 + x_i^4/24$. Here, $\kappa_i\=\omega_i^2$ follows the liquid-like VDoS $g_0(\omega)\=2\omega/\omega_0^2$ and the amplitude of anharmonicity is the same for all oscillators. The oscillators interact among themselves and with the surrounding material, especially as the liquid is quickly cooled/quenched through the glass transition and long-range elasticity builds up. The  interactions are random, reflecting the structural disorder in the emerging glass. Moreover, the quench self-organization also gives rise to internal stresses, reflecting glassy frustration, which would tend to displace the oscillators from their reference position.

At the mean-field level, i.e.~assuming each oscillator interacts with all of the others, the above physical picture corresponds to the following Hamiltonian~\cite{Kuhn_Horstmann_prl_1997,Gurevich2003,scipost_mean_field_qles_2021,meanfield_qle_pierfrancesco_prb_2021}
\begin{equation}
\label{eq:Hamiltonian}
    H = \frac{1}{2} \sum_i \kappa_i x_i^2 + \frac{1}{24}\sum_i x_i^4 + \sum_{i<j}J_{ij}x_i x_j - h \sum_i x_i \ ,
\end{equation}
of $N$ interacting anharmonic oscillators, with $i\=1\!-\!N$. Here, $J_{ij}$ are Gaussian, independent and identically distributed random variables of variance $J^2/N$ for $i\!\ne\!j$, representing random bilinear interactions between the oscillators due to structural disorder. $h$ represents internal stresses, which also emerge due to structural disorder, though $h$ itself is taken to be the same for all oscillators. $\kappa_i$, as explained above, is drawn from a rather ``featureless'' liquid-like probability distribution. The minimization of the Hamiltonian in Eq.~\eqref{eq:Hamiltonian} mimics the quench self-organization process, upon which the oscillators experience displacements and new frequencies $\omega$ at the attained minima. The statistics of the latter represent the physical nonphononic VDoS ${\cal D}_{\rm G}(\omega)$.

By studying the statistics of energy minima of Eq.~\eqref{eq:Hamiltonian} for many realizations of the disorder, $\kappa_i$ and $J_{ij}$, one obtains the resulting VDoS ${\cal D}_{\rm G}(\omega)$, which represents the VDoS of low-frequency quasi-localized vibrations in glasses. As such, the minimization of Eq.~\eqref{eq:Hamiltonian} is viewed as a procedure that transforms a ``featureless'' liquid-like VDoS $g_0(\omega)$ into a glassy VDoS ${\cal D}_{\rm G}(\omega)$. In Fig.~\ref{fig:fig2}a, we plot the initial liquid-like VDoS $g_0(\omega)$ (thin dashed line) along with a representative example of the resulting VDoS ${\cal D}_{\rm G}(\omega)$ (thick solid line). The latter features a peak at $\omega_{\rm p}$ of magnitude ${\cal D}_{\rm G}(\omega_{\rm p})$, in qualitative agreement with the experimental and simulational observations presented in Fig.~\ref{fig:fig1}. Next, we aim at deriving scaling relations for $\omega_{\rm p}$ and ${\cal D}_{\rm G}(\omega_{\rm p})$, along with the corresponding properties of the prefactor $A_{\rm g}$ of the universal $\sim\!\omega^4$ tail, also observed in Fig.~\ref{fig:fig1}.

Recently~\cite{scipost_mean_field_qles_2021}, it was shown that $A_{\rm g}\!\sim\!\exp(-c_{\rm g}\kappa_0\, h^{2/3}/J^2)$ (with $c_{\rm g}\!\simeq\!0.2$), for $y\!\equiv\!J/(h^{1/3}\kappa_0^{1/2})\!\ll\!1$. This exponential variation of $A_{\rm g}$ with $-y^{-2}$ is very reminiscent of the exponential variation of $A_{\rm g}$ with $-1/T_{\rm p}$~\cite{pinching_pnas}, where $T_{\rm p}$ is the temperature at which a supercooled liquid falls out of equilibrium during a quench, i.e.~it determines the degree of supercooling. The correspondence between the two exponential variations of $A_{\rm g}$ supports the physical relevance of the model and indicates that a decreasing $y$ implies less disordered glassy states. Moreover, the strong depletion of nonphononic vibrations in the ${\cal D}_{\rm G}(\omega)\!\sim\!\omega^4$ tail with decreasing $y$ is similar to the strong reduction in the tail of ${\cal D}_{\rm G}(\omega)$ with thermal annealing, experimentally observed in Fig.~\ref{fig:fig1}c. Consequently, we next focus on the variation of $\omega_{\rm p}$ and ${\cal D}_{\rm G}(\omega_{\rm p})$ with the disorder parameters, in the same physically relevant regime of $y\!\ll\!1$.

To derive scaling relations for $\omega_{\rm p}$ and ${\cal D}_{\rm G}(\omega_{\rm p})$ in the small $y$ regime, we need to better understand how ${\cal D}_{\rm G}(\omega)$ emerges from $g_0(\omega)$. First, we note that there exists a frequency scale $\omega_{\times}\!\sim\!h^{1/3}$ such that liquid-like vibrations with $\omega\!>\!\omega_\times$ are weakly affected by the disorder parameters $h$ and $J$, while for $\omega\!<\!\omega_\times$ liquid-like vibrations undergo significant modification (``reconstruction''), as shown in Fig.~\ref{fig:fig2}a. Second, the vast majority of the reconstructed vibrations are added on top of $g_0(\omega)$ in the frequency range $[\omega_{\rm g},\,\omega_{\times}]$ (marked by light blue-shading in Fig.~\ref{fig:fig2}a), where $\omega_{\rm g}$ is also marked therein. These vibrations constitute the peak at $\omega_{\rm p}$. A small fraction of the reconstructed vibrations populate the frequency range $[0\,,\omega_{\rm g}]$ (marked by light green-shading in Fig.~\ref{fig:fig2}a), including those in the universal $\sim\!\omega^4$ tail.

\begin{figure*}[ht!]
\center
\includegraphics[width=0.85\textwidth]{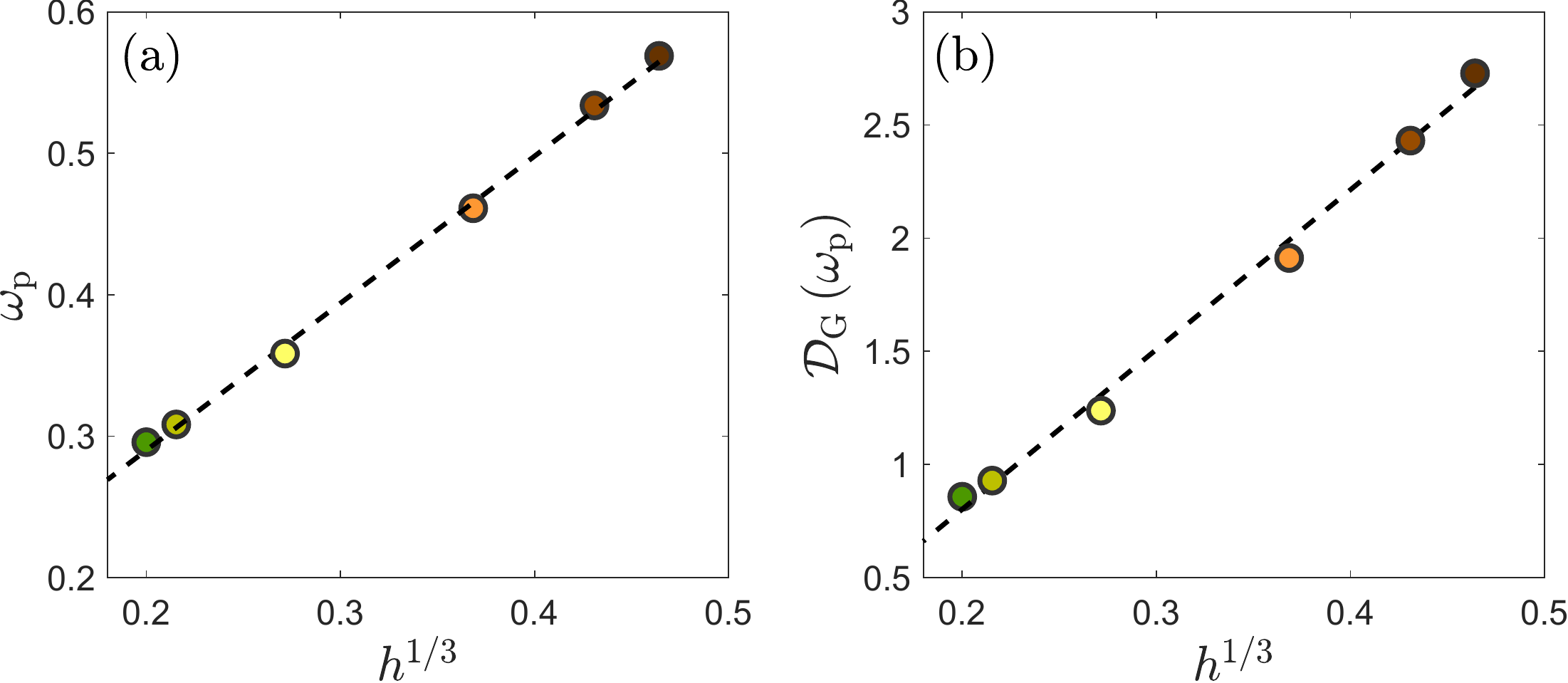}% Here is how to import EPS art
\caption{(a) The prediction $\omega_{\rm p}\!\sim\!h^{1/3}$ of Eq.~\eqref{eq:BP_scaling} is verified using numerical solutions of the mean-field model with $J\!=\!0.05$ and $h\!=\![0.1,0.08,0.05,0.02,0.01,0.008]$ (recall that $\kappa_0\!=\!1$), corresponding to small $y\!=\!J/(h^{1/3}\kappa_0^{1/2})$ values (between $\sim\!0.11$ to $0.25$). The color corresponds to the different $h$ values. (b) The prediction ${\cal D}_{\rm G}(\omega_{\rm p})\!\sim\!h^{1/3}$ of Eq.~\eqref{eq:BP_scaling} is verified using the same numerical solutions of the mean-field model used in panel (a). The dashed lines are guides to the eye. Note that while $\omega_{\rm p}$ and ${\cal D}_{\rm G}(\omega_{\rm p})$ increase by approximately a factor of $2$ over the presented range of disorder, $A_{\rm g}$ decreases by many  orders of magnitude (not shown).}
\label{fig:fig3}
\end{figure*}

The number of vibrations in the frequency range $[0,\,\omega_{\times}]$ is conserved upon reconstruction. This, together with neglecting the number of reconstructed vibrations in $[0\,,\omega_{\rm g}]$ compared to those that populate the peak region in $[\omega_{\rm g},\,\omega_{\times}]$, yield (SM S-1C)
\begin{equation}
\omega_{\rm p}\sim h^{1/3}\qquad\hbox{and}\qquad {\cal D}(\omega_{\rm p})\sim h^{1/3}/\omega_0^2 \ .
\label{eq:BP_scaling}
\end{equation}
The scaling predictions in Eq.~\eqref{eq:BP_scaling} suggest that while $J\!>\!0$ (accounting for interactions between vibrations) is essential for the emergence of the universal $\sim\!\omega^4$ tail (recall that $A_{\rm g}\!\to\!0$ as $J\!\to\!0$), it contributes only to sub-leading orders in the peak properties, in the $y\=J/(h^{1/3}\kappa_0^{1/2})\!\ll\!1$ limit of interest. In Fig.~\ref{fig:fig3}, the scaling predictions in Eq.~\eqref{eq:BP_scaling} are quantitatively verified by numerical solutions of the mean-field model. Importantly, Eq.~\eqref{eq:BP_scaling}, along with $A_{\rm g}\!\sim\!\exp(-c_{\rm g}\kappa_0\, h^{2/3}/J^2)$, indicate that an increase in the disorder parameter $h$ at fixed $J$ gives rise to strong (exponential) reduction in the $\sim\!\omega^4$ tail, but to weak (power-law) increase in $\omega_{\rm p}$ and ${\cal D}_{\rm G}(\omega_{\rm p})$. These predictions fully agree with the experimental trends presented in Fig.~\ref{fig:fig1}.

%%%%%%%%%%%%%%%%%%%%%%%%%%%%%%%%%%%%%%%%%%%%%%%%
%%%%%%%%%%%%%%%%%%%%%%%%%%%%%%%%%%%%%%%%%%%

\section{The nature of the boson peak modes and their localization properties}

Up to now, we focused on the VDoS ${\cal D}_{\rm G}(\omega)$ and its properties, but not on the nature of the boson peak vibrational modes themselves, in particular their spatial structure. Current experimental techniques entirely lack the spatial resolution to address this issue, and hence we approach it in the framework of the mean-field model and atomistic computer simulations. In the context of the former, the question boils down to quantifying how many oscillators are taking part in each vibration at minima of the Hamiltonian, i.e.~the degree of localization of vibrations, commonly quantified through the averaged participation ratio $\bar{e}(\omega)$ (SM S-1D). A vibration that is fully localized at a single oscillator features a participation ratio of $1/N$ (recall that $N$ is the total number of oscillators). On the other hand, a fully delocalized vibration features a participation ratio of unity.

In Fig.~\ref{fig:fig2}b, we present $\bar{e}(\omega)$, corresponding to ${\cal D}_{\rm G}(\omega)$ shown in Fig.~\ref{fig:fig2}a. $\bar{e}(\omega)$ attains a peak near $\omega_{\rm p}$, i.e.~$\bar{e}(\omega)$ is peaked in the boson peak region~\cite{Gurevich2005}. In addition, Fig.~\ref{fig:fig2}b shows that $N\bar{e}(\omega_{\rm p})\!\gg\!1$, i.e.~it suggests that boson peak vibrational modes feature many coupled quasi-localized vibrations, while modes in the $\sim\!\omega^4$ tail feature $N\bar{e}(\omega_{\rm p})$ that is orders of magnitude smaller~\cite{modes_prl_2016}. Finally, $\bar{e}(\omega)$ plateaus at frequencies above the peak.
%%%%%%%%%%%%%%%%%%%%%%%%%%%%%%%%%%%%%%%%%%%%%
\begin{figure*}[ht!]
\includegraphics[width=1\textwidth]{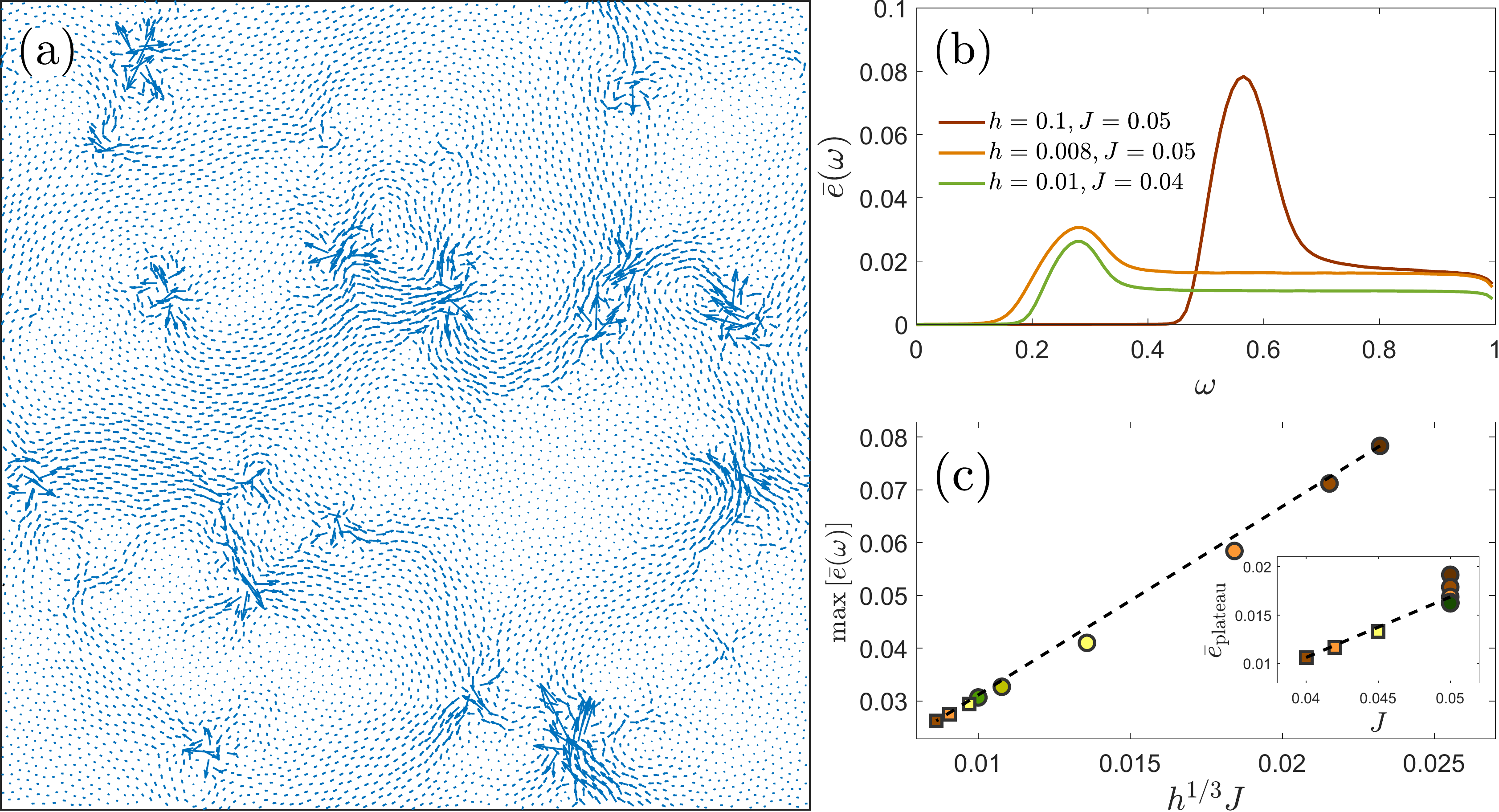}% Here is how to import EPS art
\caption{(a) A superposition of quasi-localized nonphononic vibrations identified inside a boson peak vibrational model in the same canonical computer glass model (in two dimensions) used in Fig.~\ref{fig:fig1}, see~\cite{boson_peak_2d_arXiv2022} for details. (b) The averaged participation ratio $\bar{e}(\omega)$ for a few sets of the disorder parameters (see legend). (c) The prediction $\mbox{max}[\bar{e}(\omega)]\!\sim\!h^{1/3}J$ (see text for details) is verified using numerical solutions of the mean-field model with $J\!=\!0.05$ and $h\!=\![0.1,0.08,0.05,0.02,0.01,0.008]$ (circles, the same as in Fig.~\ref{fig:fig3}), and with $h\!=\!0.01$ with $J\!=\![0.04,0.042,0.045]$ (squares), all corresponding to small $y\!=\!J/(h^{1/3}\kappa_0^{1/2})$ values. $\mbox{max}[\bar{e}(\omega)]$ is the maximum of $\bar{e}(\omega)$, attained very close to $\omega_{\rm p}$ (cf.~Fig.~\ref{fig:fig2}). (inset) The prediction $\bar{e}_{\text{plateau}}\!\sim\!J$ (see text for details) is verified using the same numerical solutions as in the main panel (same symbols and colors). $\bar{e}_{\text{plateau}}$ corresponds to the plateau value of $\bar{e}(\omega)$, above the maximum. The dashed lines are guides to the eye.}
\label{fig:fig4}
\end{figure*}
%%%%%%%%%%%%%%%%%%%%%%%%%%%%%%%%%%%%%%%%%%%%%

The prediction that boson peak vibrational modes feature many coupled quasi-localized vibrations poses serious challenges. Testing it in computer glasses requires tools for identifying quasi-localized vibrations inside boson peak vibrational modes, possibly featuring many hybridized/mixed quasi-localized vibrations and extended phonons~\cite{Schober_Oligschleger_numerics_PRB,Gurevich2003,Gurevich2007,boson_peak_2d_arXiv2022}. Such tools have just began to emerge~\cite{JCP_Perspective,boson_peak_2d_arXiv2022}, and Fig.~\ref{fig:fig4}a presents a preliminary example (corresponding to Fig.~3d in~\cite{boson_peak_2d_arXiv2022}). It shows a superposition of identified quasi-localized vibrations in a boson peak vibrational mode in a two-dimensional computer glass (the full boson peak vibrational mode is shown in Fig.~3c in~\cite{boson_peak_2d_arXiv2022}), clearly supporting the prediction of the existence of many coupled quasi-localized vibrations.

We next derive scaling relations for $\bar{e}(\omega)$, shown in Fig.~\ref{fig:fig2}b, in the framework of the mean-field model. In Fig.~\ref{fig:fig4}b, we present $\bar{e}(\omega)$ for several combinations of the disorder parameters $J$ and $h$ (we set $\kappa_0\=1$, as done elsewhere in this work), all in the $y\=J/(h^{1/3}\kappa_0^{1/2})\!\ll\!1$ regime of interest. In the $\sim\!\omega^4$ tail region, vibrations in the mean-field model are rather localized~\cite{meanfield_qle_pierfrancesco_prb_2021}, corresponding to the very small observed values of $\bar{e}(\omega\!\to\!0)$ (cf.~Fig.~\ref{fig:fig4}b). We do not discuss the tail region here, but rather focus on the peak region and the plateau that follows it.

$\bar{e}(\omega)$ emerges from interactions between the oscillators, mediated by the bilinear coupling coefficients $J_{ij}$, i.e.~one trivially has $N\bar{e}(\omega)\=1$ for $J\=0$. Moreover, since we consider the $y\=J/(h^{1/3}\kappa_0^{1/2})\!\ll\!1$ regime (and set $\kappa_0\=1$), we can treat the effect of weak interactions of characteristic size $J$ as perturbations on top of the $J\=0$ and $h\!>\!0$ case. The latter is fully described by the single-oscillator Hamiltonian $\kappa x^2/2 + x^4/24-hx$, implying that liquid-like vibrations below $\kappa\!\sim\!h^{2/3}$ are blue-shifted and accumulate in a narrow stiffness range near $\kappa\!\sim\!h^{2/3}$, while those above $\kappa\!\sim\!h^{2/3}$ are weakly affected (SM S-1D). When weak random interactions are introduced, $J\!>\!0$, the number of oscillators that contribute to reconstructed vibrations must scale with $J$ to leading order. Near $\omega_{\rm p}\!\sim\!h^{1/3}$, there are $\sim\!Nh^{2/3}$ blue-shifted vibrations and hence the participation ratio after reconstruction satisfies $N\bar{e}(\omega_{\rm p})\!\sim\!\sqrt{N}h^{1/3}J$, implying $\mbox{max}[\bar{e}(\omega)]\!\sim\!h^{1/3}J/\sqrt{N}$.

This scaling prediction is verified in Fig.~\ref{fig:fig4}c, showing that internal stresses, represented by $h$, and disorder-mediated interactions between the oscillators, represented by the standard deviation $J/\sqrt{N}$ of the random variables $J_{ij}$, lead to $N\bar{e}(\omega)\!\gg\!1$ near the boson peak. That is, the mean-field model predicts that boson peak modes feature many coupled quasi-localized vibrations, as recently observed in computer glasses (cf.~Fig.~\ref{fig:fig4}a). Moreover, note that $\bar{e}(\omega_{\rm p})\!\sim\!h^{1/3}J$, together with Eq.~\eqref{eq:BP_scaling} (verified in Fig.~\ref{fig:fig3}), implies $\bar{e}(\omega_{\rm p})\!\sim\!{\cal D}(\omega_{\rm p})J$. Finally, above $\omega_{\rm p}\!\sim\!h^{1/3}$, the stiffnesses are uniformly distributed and weakly affected by $h$, hence we have $\bar{e}_{\rm plateau}\!\sim\!J/\sqrt{N}$, where $\bar{e}_{\rm plateau}$ is the plateau level of $\bar{e}(\omega)$ (cf.~Fig.~\ref{fig:fig4}b). This prediction is verified in the inset of Fig.~\ref{fig:fig4}c.

\section{Conclusions and outlook}

Our results shed basic light on the origin, nature and properties of the universally-observed boson peak in glasses. We followed~\cite{surovtsev2003density} and showed that the nonphononic part of the VDoS ${\cal D}_{\rm G}(\omega)$ features an intrinsic peak at $\omega_{\rm p}$, which is distinguished from the conventionally defined peak in the reduced VDoS ${\cal D}(\omega)/\omega^2$, commonly denoted by $\omega_{_{\rm BP}}$. While the definition of $\omega_{_{\rm BP}}$ involves Debye's VDoS of phonons, $\omega_{\rm p}$ is an intrinsic property of the nonphononic VDoS. Moreover, our results indicate --- at the fundamental ontological level --- that the excess vibrations that constitute the boson peak are the very same quasi-localized nonphononic vibrations that populate the universal $\sim\!\omega^4$ tail of ${\cal D}_{\rm G}(\omega)$.

It is shown that vibrational modes near the boson peak feature many more coupled quasi-localized nonphononic vibrations than modes in the universal tail. Moreover, the peak frequency $\omega_{\rm p}$ and its magnitude ${\cal D}_{\rm G}(\omega_{\rm p})$ mildly increase upon thermal annealing (i.e.~with decreasing degree of glassy disorder), while ${\cal D}_{\rm G}(\omega)$ is strongly reduced in the tail region under the same conditions. Future work, both experimental and simulational, should further test these predictions for a broader range of glasses under different nonequilibrium histories. Future work should also explore the implications of our findings to other glass properties, such as the specific heat, not discussed here. Finally, all of our findings are semi-quantitatively explained by the solution of a mean-field model of interacting quasi-localized vibrations, which also predicts the universal $\sim\!\omega^4$ tail of ${\cal D}_{\rm G}(\omega)$. As such, future work should further explore the predictive powers of the model in relation to other properties of glasses.\\

{\bf \large Acknowledgments} A.M.~acknowledges support from the Minerva center on ``Aging, from physical materials to human tissues''. E.L.~acknowledges support from the NWO (Vidi grant no.~680-47-554/3259). E.B.~acknowledges support from the Ben May Center for Chemical Theory and Computation and the Harold Perlman Family.

\newpage

%%%%%%%%%%%%%%%%%%%%%%%%%%%%%%%%%%%%%%%%%%%%%%%%%%%%%%%%%%%%%%%%%%%%%%%%%%%%%%%%%
%%%%%%%%%%%%%%%%%%%%%% these lines of code handle the concatenation %%%%%%%%%%%%%
%%%%%%%%%%%%%%%%%%%%%%%%%%%%%%%%%%%%%%%%%%%%%%%%%%%%%%%%%%%%%%%%%%%%%%%%%%%%%%%%%

\onecolumngrid
\newpage
\begin{center}
\textbf{\large Supplementary Materials for\\``The boson peak in the vibrational spectra of glasses''}
\end{center}
\twocolumngrid
\setcounter{equation}{0}
\setcounter{figure}{0}
\setcounter{table}{0}
\setcounter{section}{0}
\setcounter{page}{1}
\makeatletter
\renewcommand{\theequation}{S\arabic{equation}}
\renewcommand{\thesection}{S-\Roman{section}}
\renewcommand{\thefigure}{S\arabic{figure}}
\renewcommand*{\thepage}{S\arabic{page}}
%\renewcommand{\bibnumfmt}[1]{[S#1]}
%\renewcommand{\citenumfont}[1]{S#1}
%%%%%%%%%%%%%%%%%%%%%%%%%%%%%%%%%%%%%%%%%%%%%%%%%%%%%%%%%%%%%%%%%%%%%%%%%%%%%%%%%
%%%%%%%%%%%%%%%%%%%%%% these lines of code handle the concatenation %%%%%%%%%%%%%
%%%%%%%%%%%%%%%%%%%%%%%%%%%%%%%%%%%%%%%%%%%%%%%%%%%%%%%%%%%%%%%%%%%%%%%%%%%%%%%%%

\section{Methods}

\subsection{Reanalysis of thermal annealing\\ experimental data}

\begin{figure}[ht!]
\center
\includegraphics[width=0.48\textwidth]{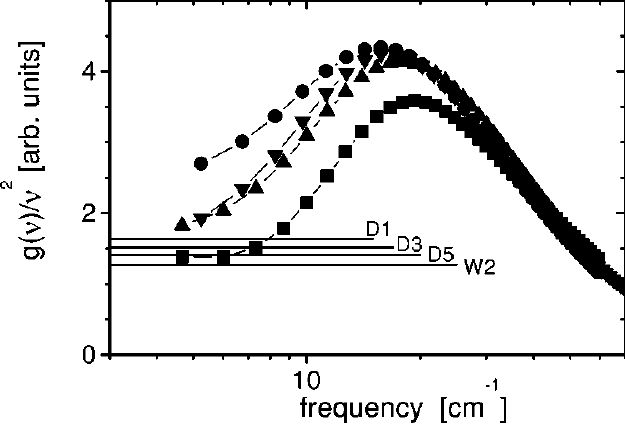}
\caption{The experimental data of~\cite{surovtsev2003density}, adapted as is from Fig.~3 therein. See text for extensive discussion. Note that the variation with annealing of the properties of the peak --- i.e.~its frequency $\nu_{_{\rm BP}}$ and magnitude --- in the conventional reduced-VDoS presentation is just opposite to the variation shown in Fig.~1b in the manuscript, using the nonphononic VDoS ${\cal D}_{\rm G}(\nu)$. Similar qualitative differences associated with the two presentations were highlighted in~\cite{KALAMPOUNIAS20064619,ruocco_boson_peak_ruocco_2006}.}
\label{fig:figS1}
\end{figure}

The data presented in Figs.~1b-c in the manuscript correspond to the experimental data digitized from Fig.~3 in~\cite{surovtsev2003density}, reproduced here in its original form in Fig.~\ref{fig:figS1}, and to the values of $A_{\rm D}$ appearing therein. In Fig.~\ref{fig:figS1} (i.e.~Fig.~3 in~\cite{surovtsev2003density}), the measured VDoS $g(\nu)$ (denoted by ${\cal D}(\nu)$ in the manuscript and obtained using low-frequency Raman scattering) divided by $\nu^2$ was plotted  ($\nu$ is the Raman wavenummber shift in units of cm$^{-1}$) for boron-oxide (B$_2$O$_3$) glass samples of different thermal histories. The different samples were given names and the corresponding data were denoted by different symbols, as indicated in Fig.~\ref{fig:figS1}. The names and symbols, from the top curve to the bottom one are: `D1' and circles, `D3' and down-triangles, `D5' and up-triangles, and `W2' and squares. The thermal history of each sample is detailed in Table~\ref{tab:exp}.

The `D1' sample is termed `As quenched' in the manuscript, `D3' is termed `Annealed I', `D5' is termed `Annealed II', and `W2' is termed `Most annealed'. We also follow the same symbols scheme in Figs.~1b-c in the manuscript, except that we replaced the down-triangles for the `D3' sample data by diamonds for improved visual clarity (as well as colors). We digitized the data for $g(\nu)/\nu^2$ shown in Fig.~\ref{fig:figS1} and multiplied it by $\nu^2$ to obtain ${\cal D}(\nu)\=g(\nu)$. The horizontal solid lines in Fig.~\ref{fig:figS1} correspond to the values of $A_{\rm D}$ for each sample, which we digitized as well; the values are also reported in Table~\ref{tab:exp}. We then plotted the nonphononic VDoS ${\cal D}_{\rm G}(\nu)\!\equiv\!{\cal D}(\nu)\!-\!A_{\rm D}\,\nu^2$ in Figs.~1b-c in the manuscript.

The position of the peak of ${\cal D}_{\rm G}(\nu)$, denoted by $\nu_{\rm p}$, is reported in Table~\ref{tab:exp} for the different thermal histories, as well as the peak value, ${\cal D}_{\rm G}(\nu_{\rm p})$. For completeness, we added in Table~\ref{tab:exp} the mass density $\rho$ of each sample (extracted from Table 1 of~\cite{surovtsev2003density}). Moreover, we used Debye's velocity values $v_{\rm d}$ from~\cite{surovtsev2003density}, together with the longitudinal (dilatational) wave-speed $v_{\rm l}$ values from~\cite{surovtsev2000inelastic}, to extract the transverse (shear) wave-speed $v_{\rm t}$ using the relation $3v_{\rm d}^{-3}\=v_{\rm l}^{-3} +2v_{\rm t}^{-3}$~\cite{surovtsev2000inelastic}. Then, we used the density $\rho$ to extract the shear modulus $\mu\=\rho\,v_{\rm t}^2$, as reported in Table~\ref{tab:exp}. The obtained $\mu$ values are consistent with those of~\cite{ramos1997correlation}. Finally, to highlight the variability of all physical quantities with annealing, we added to each column in Table~\ref{tab:exp} the relative variation with respect to the as quenched sample (in angular brackets).

%%%%%%%%%%%%%%%%%%%%%%%%%%%%%%%%%%%%%%%%%%%%%%%%%%%%
%%%%%%%%%%%%%%%%%%%%%%%%%%%%%%%%%%%%%%%%%%%%%%%%%%%%

%%%%%%%%%%%%%%%%%%%%%%%%%%%%%%%%%%%%%%%
\begin{table*}[ht!]
    \centering
    \begin{tabular}{|c||c|c|c|c|c|c|}
    \hline
         Symbol &
          Thermal history &
          $A_{\rm D}$ [a.u.] &
          $\nu_{\rm p}\,[\text{cm}^{-1}]$  &
          ${\cal D}_{\rm G}(\nu_{\rm p})$\,[a.u.] &
          $\rho$  [g/cm$^3$] &
          $\mu$ [GPa] \\
        \hline    \hline
         D1 (As quenched) & As quenched & 1.639 (1) & 30.02 (1) &  1365.91 (1) & 1.804 (1)  & 6.14 (1)  \\
         D3 (Annealed I) & 530 K, 50 h & 1.514 (0.92) & 30.37 (1.01) &  1529.48 (1.12) & 1.826 (1.01) &  6.52 (1.06) \\
         D5 (Annealed II) & 480 K, 170 h & 1.417 (0.86) & 31.44 (1.05) & 1563.07 (1.14) & 1.834 (1.02) &  6.89 (1.12) \\
         W2 (Most annealed) & 490 K, 100 h &1.292 (0.79) &  36.16 (1.20) &  1520.80 (1.11) &  1.866 (1.03) & 7.50 (1.22) \\
    \hline
    \end{tabular}
    \caption{The thermal history (annealing treatment) of the different boron-oxide (B$_2$O$_3$) glass samples of~\cite{surovtsev2003density,surovtsev2000inelastic}, and all physical quantities that characterize them. The thermal history, except for the as quenched sample, is characterized by the annealing temperature (in Kelvin) and annealing time (in hours) applied to the as quenched sample, see second column. $A_{\rm D}$, $\rho$ and $\mu$ are extracted from the existing literature (see text for details), while $\nu_{\rm p}$ and ${\cal D}(\nu_{\rm p})$ are obtained from Fig.~1b in the manuscript. The relative variation with annealing of each quantity, relative to the as quenched sample (`D1') is reported in angular brackets in each column.}
    \label{tab:exp}
\end{table*}
%%%%%%%%%%%%%%%%%%%%%%%%%%%%%%%%%%%%%%%

\subsection{Atomistic computer glasses}

We employ a simple glass-forming model in three dimensions~\cite{cge_paper} in which half of the particles are `large' and half are `small'. The particles of both species have equal mass $m$. The pairwise potential of this model is given by
\begin{equation}    \varphi(r,\lambda)/\varepsilon = \left(\frac{\lambda}{r}\right)^{10} + \sum_{\ell=0}^3c_{2\ell}\left(\frac{r}{\lambda}\right)^{2\ell}\,,
\label{Seq:potential}
\end{equation}
where $\varepsilon$ denotes our microscopic units of energy, $r$ is the pairwise distance between two particles, $c_{2\ell}$ are coefficients (reported in Table~\ref{tab:c_2ell} below) that guarantee the smoothness of the potential at the dimensionless cutoff distance $r_{\rm c}\!=\!1.48\lambda$, where  $\lambda\!=\!1.4\lambdabar$ for `large'-`large' pairs, $\lambda\!=\!1.18\lambdabar$ for `large'-`small' pairs and $\lambda\!=\!1.0\lambdabar$ for `small'-`small' pairs. $\lambdabar$ denotes the microscopic units of length. We fix the number density at $N/V\!=\!0.82$ for all simulations ($N$ denotes the total number of particles and $V$ denotes the volume). All dimensional observables reported below and in the manuscript should be understood as expressed in terms of the aforementioned microscopic units $m,\varepsilon$ and $\lambdabar$.

%%%%%%%%%%%%%%%%%%%%%%%%%%%%%%%%%%%%%%%
\begin{table}[ht!]
    \centering
    \begin{tabular}{|c|c|}
    \hline
         $c_0$ & -1.1106337662511798 \\ \hline
         $c_2$ & 1.2676152372297065 \\ \hline
         $c_4$ & -0.4960406072849212 \\ \hline
         $c_6$ & 0.0660511826415732 \\ \hline
    \end{tabular}
    \caption{The coefficients $c_{2\ell}$ appearing in Eq.~\eqref{Seq:potential}.}
    \label{tab:c_2ell}
\end{table}
%%%%%%%%%%%%%%%%%%%%%%%%%%%%%%%%%%%%%%%

We prepared glassy samples following two protocols; for both protocols, we first equilibrate high temperature liquid states at $T\!=\!1.0$. Then, in the first protocol, we perform an instantaneous quench (hyperquench) using a standard conjugate gradient minimization algorithm, corresponding to an infinite quench rate $\dot{T}\!\to\!\infty$. In the second protocol, we cool the system at a finite cooling rate $\dot{T}\!=\!10^{-3}$, removing any remnant heat deep in the glass phase with a potential-energy minimization. Using these two glass-formation protocols, we prepared 40 independent realizations of $N\!=\!4\!\times\!10^6$ particles.

\subsubsection*{VDoS calculations}

The VDoS of large computer glasses of several millions of particles can be obtained using the \emph{Kernel Polynomial Method} (KPM)~\cite{kpm_review_2006}. We followed exactly the procedure as described in detail in~\cite{Tanguy_2016}. The KPM requires choosing the \emph{truncation degree} $K$, and the number $R$ of initial random vectors used in the calculation. We chose $K\!=\!3000$ and $R\!=\!10$, in addition to carrying out this analysis over 40 independent glasses.

%\newpage

\subsection{The mean-field model: \\Numerical solutions and scaling relations for ${\cal D}_{\rm G}(\omega)$}

The main properties of the nonphononic VDoS ${\cal D}_{\rm G}(\omega)$ in the framework of the mean-field model, defined in Eq.~(1) in the manuscript (see details provided therein), are extensively discussed in the manuscript. Here, we provide details of the numerical solution procedure and some supporting results in relation to the scaling predictions discussed in the manuscript.

Numerical solutions for the statistics of energy minima of the Hamiltonian in Eq.~(1) in the manuscript are obtained as follows. We initiated $M\=2000$ different realizations of $N\!=\!16000$ coupled oscillators each. The initial oscillators' positions $x_i^{\mbox{\tiny{(0)}}}$ were set randomly in the range $x_i^{\mbox{\tiny{(0)}}}\in\left[-0.005,0.005\right]$. These initial positions generate non-vanishing net forces on the oscillators. After initiation, we used a gradient descent algorithm to relax the oscillators to the closest mechanically-stable energy minimum, resulting in displacements $x_i$. Following this minimization procedure, we calculated the Hessian matrix $\mathcal{M}_{ij}\!\equiv\!\tfrac{\partial ^2 H}{\partial x_i \partial x_j}$ at this newly-attained energy minima, and diagonalized it to find the eigenmodes ${\bm \psi}$ and their corresponding eigenvalues $\omega^2$, according to ${\calBold  M}\!\cdot\! {\bm \psi}\=\omega^2 {\bm \psi}$ (${\bm \psi}$ is normalized,  $\sum_i |\psi_i|^2\=1$). ${\cal D}_{\rm G}(\omega)$ is obtained by generating a histogram over the vibrational frequencies collected from all realizations, as presented in Fig.~2a in the manuscript. This procedure was repeated for every pair of $J$ and $h$ values.

As we exclusively focused on the $y\!\equiv\!J/(h^{1/3}\kappa_0^{1/2})\!\ll\!1$ regime, we followed~\cite{scipost_mean_field_qles_2021} and treated the characteristic interactions strength $J$ as a small perturbation on top of the characteristic internal force $h$. It was shown in~\cite{scipost_mean_field_qles_2021} that there exists a characteristic frequency scale $\omega_{\times}$ that splits the entire frequency domain $[0, \omega_0]$ into two parts, one in which the VDoS undergoes ``reconstruction'', $0\!<\!\omega\!<\!\omega_{\times}$, and another in which it mostly does not, $\omega_{\times}\!<\!\omega\!<\!\omega_0$. The scaling prediction for $\omega_{\times}$ takes the form $\omega_{\times}\!\sim\!h^{1/3}(1+c_{\times}y)$~\cite{scipost_mean_field_qles_2021}, where the two leading orders in $y$ are included (note that in the manuscript only the leading order $\omega_{\times}\!\sim\!h^{1/3}$ is discussed).

\begin{figure}[ht!]
\center
\includegraphics[width=0.45\textwidth]{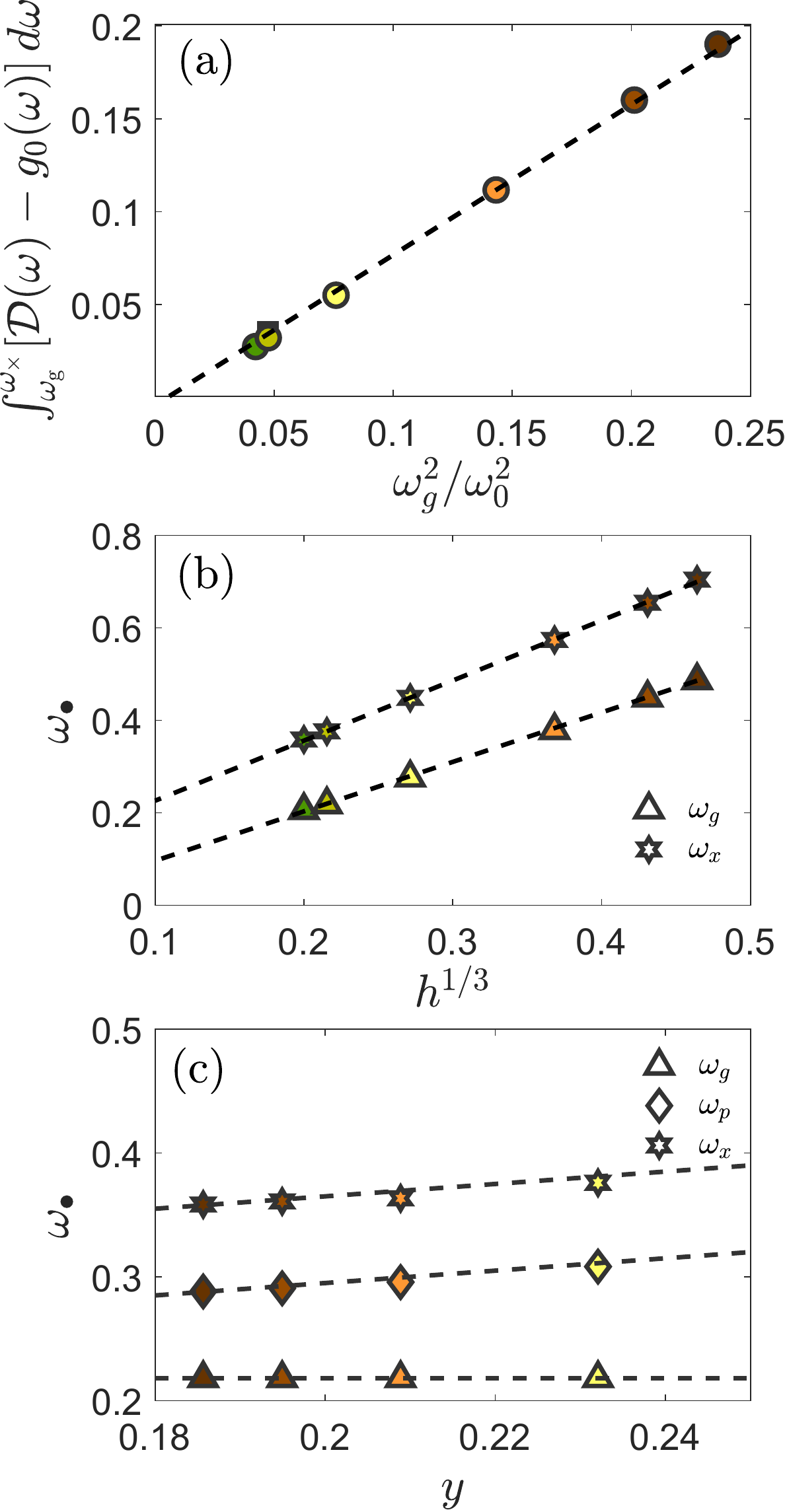}
\caption{(a) Verification of Eq.~\eqref{eq:conservation}. Circles correspond to fixed $J\!=\!0.05$ with $h\!=\![0.1,0.08,0.05,0.02,0.01,0.008]$ and squares correspond to fixed $h\!=\!0.01$ with $J\!=\!\left[0.04,0.042,0.045\right]$. We used $\omega_0\!=\!1$ as done throughout this work. (b) Verification of the $\sim\!h^{1/3}$ scaling of $\omega_{\rm g}$ and $\omega_\times$ (see text) for $J\!=\!0.05$ and $h\!=\![0.1,0.08,0.05,0.02,0.01,0.008]$. The corresponding prediction for $\omega_{\rm p}$ is verified in Fig.~3a in the manuscript. (c) Verification of the sub-leading (linear) contribution in $y$ in the prediction $\omega_\bullet\!\sim\!h^{1/3} \left(1+c_{\times}y \right)$ for $\omega_{\rm g}$, $\omega_{\rm p}$ and $\omega_\times$, for fixed $h\!=\!0.01$ with $J\!=\!\left[0.04,0.042,0.045, 0.05\right]$.
\label{fig:figS2}}
\end{figure}

Focusing on the ``reconstructed'' domain $0\!<\!\omega\!<\!\omega_{\times}$, we considered the frequency $\omega_{\rm g}$ defined according to the lowest $\omega\!>\!0$ solution to ${\cal D}_{\rm G}(\omega)\=g_0(\omega)$, see Fig.~2a in the manuscript. Modes below $\omega_{\rm g}$ are blue-shifted (by amount $\sim\!h^{1/3}$) to the frequency domain $\omega_{\rm g}\!<\!\omega\!<\!\omega_{\times}$, see~\cite{scipost_mean_field_qles_2021} and Fig.~\ref{fig:figS3}. A tiny fraction of these blue-shifted modes are ``pushed back'' to the frequency domain $0\!<\!\omega\!<\!\omega_{\rm g}$ by interaction-induced fluctuations, forming the gapless (or pseudo-gapped) ${\cal D}_{\rm G}(\omega)\=A_{\rm g}\,\omega^4$ tail. The latter implies $\int_0^{\omega_{\rm g}}{\cal D}_{\rm G}(\omega)\,d\omega\!\ll\omega_{\rm g}^2/\omega_0^2$. Since the ``reconstruction'' of ${\cal D}_{\rm G}(\omega)$ in the frequency domain $0\!<\!\omega\!<\!\omega_{\times}$ only redistributes the modes, conservation of modes then implies
\begin{equation}
\label{eq:conservation}
\int_{\omega_{\rm g}}^{\omega_{\times}}\!\left[{\cal D}_{\rm G}(\omega)-g_0(\omega)\right]d\omega\simeq\omega_{\rm g}^2/\omega_0^2 \ ,
\end{equation}
which is verified in Fig.~\ref{fig:figS2}a.

%%%%%%%%%%%%%%%%%%%%%%%%%%%%%%%%%%%%%%%%%%%%%%%%%%%%
%%%%%%%%%%%%%%%%%%%%%%%%%%%%%%%%%%%%%%%%%%%%%%%%%%%%

The validity of Eq.~\eqref{eq:conservation} justifies neglecting the light green-shaded area in Fig.~2a in the manuscript compared to the light blue-shaded therein, as done in the manuscript. In addition, we estimate $\int_{\omega_{\rm g}}^{\omega_{\times}} {\cal D}_{\rm G}(\omega)\, d\omega$ as ${\cal D}_{\rm G}(\omega_\times)(\omega_\times-\omega_{\rm g})$, which implies $\omega_{\rm g}\!\sim\!\omega_\times\!\sim\!h^{1/3}$ to leading order in smallk $y$.  This prediction is verified in Fig.~\ref{fig:figS2}b. Since scaling-wise we have $\omega_{\rm p}\!\sim\!(\omega_{\rm g}+\omega_{\times})/2$, we end up with
\begin{equation}
\omega_{\rm p}\sim\omega_{\rm g}\sim\omega_{\times}\sim h^{1/3}\left(1+c_{\times}y \right) \ ,
\label{eq:omegaBulletScaling}
\end{equation}
which is verified in Figs.~\ref{fig:figS2}c. Finally, by estimating Eq.~\eqref{eq:conservation} as $\int_{\omega_{\rm g}}^{\omega_{\times}}\!\left[{\cal D}_{\rm G}(\omega)\!-\!g_0(\omega)\right]d\omega\!\simeq\!\left[{\cal D}_{\rm G}(\omega_{\rm p})\!-\!2\omega_{\rm p}/\omega_0^2\right](\omega_{\times}\!-\!\omega_{\rm g})\!\sim\!\omega_{\rm g}^2/\omega_0^2$, Eq.~(2) in the manuscript is obtained.

\subsection{The mean-field model: The average participation ratio $\bar{e}(\omega)$}

The averaged participation ratio $\bar{e}(\omega)$ and its scaling properties are extensively discussed in the manuscript. Here, we provide the relevant definition, the numerical averaging procedure and some supporting data referred to in the manuscript.

The participation ratio of an eigenmode ${\bm \psi}^{\mbox{\tiny{(j)}}}$, which is a normalized solution to ${\calBold M}\!\cdot\! {\bm \psi}^{\mbox{\tiny{(j)}}}\=[\omega^{\mbox{\tiny{(j)}}}]^2\, {\bm \psi}^{\mbox{\tiny{(j)}}}$ (i.e.~$\sum_i |\psi_i^{\mbox{\tiny{(j)}}}|^2\=1$), is defined as
\begin{equation}
e^{\mbox{\tiny{(j)}}} \equiv \frac{1}{N \sum_{i=1}^N \left[\psi_i^{\mbox{\tiny{(j)}}}\right]^4} \,
\end{equation}
where the index $i$ corresponds to the projection on the $i^{\mbox{\tiny{th}}}$ oscillator. Finding the complete set of eigenmodes ${\bm \psi}^{\mbox{\tiny{(j)}}}$ per realization of the disorder (with fixed $J$, $h$ and $\kappa_0$), we sort $e^{\mbox{\tiny{(j)}}}$ according to their corresponding frequencies $\omega^{\mbox{\tiny{(j)}}}$, collect data from all $M$ realizations and average over bins of size $\Delta \omega$ to obtain $\bar{e}(\omega)$. That is, we define
\begin{equation}
    \bar{e}(\omega) \equiv \langle e(\omega)\rangle_{\omega, \omega+\Delta \omega} \ ,
\end{equation}
where $\langle\bullet\rangle_{\omega, \omega+\Delta \omega}$ denotes averaging over modes with frequencies between $\left[\omega, \omega+\Delta \omega\right]$. In particular, we used $\Delta\omega\=0.01$ to produce the averaged participation ratio curves in Fig.~2b and Fig.~4 in the manuscript.

\begin{figure}[ht!]
\center
\includegraphics[width=0.4\textwidth]{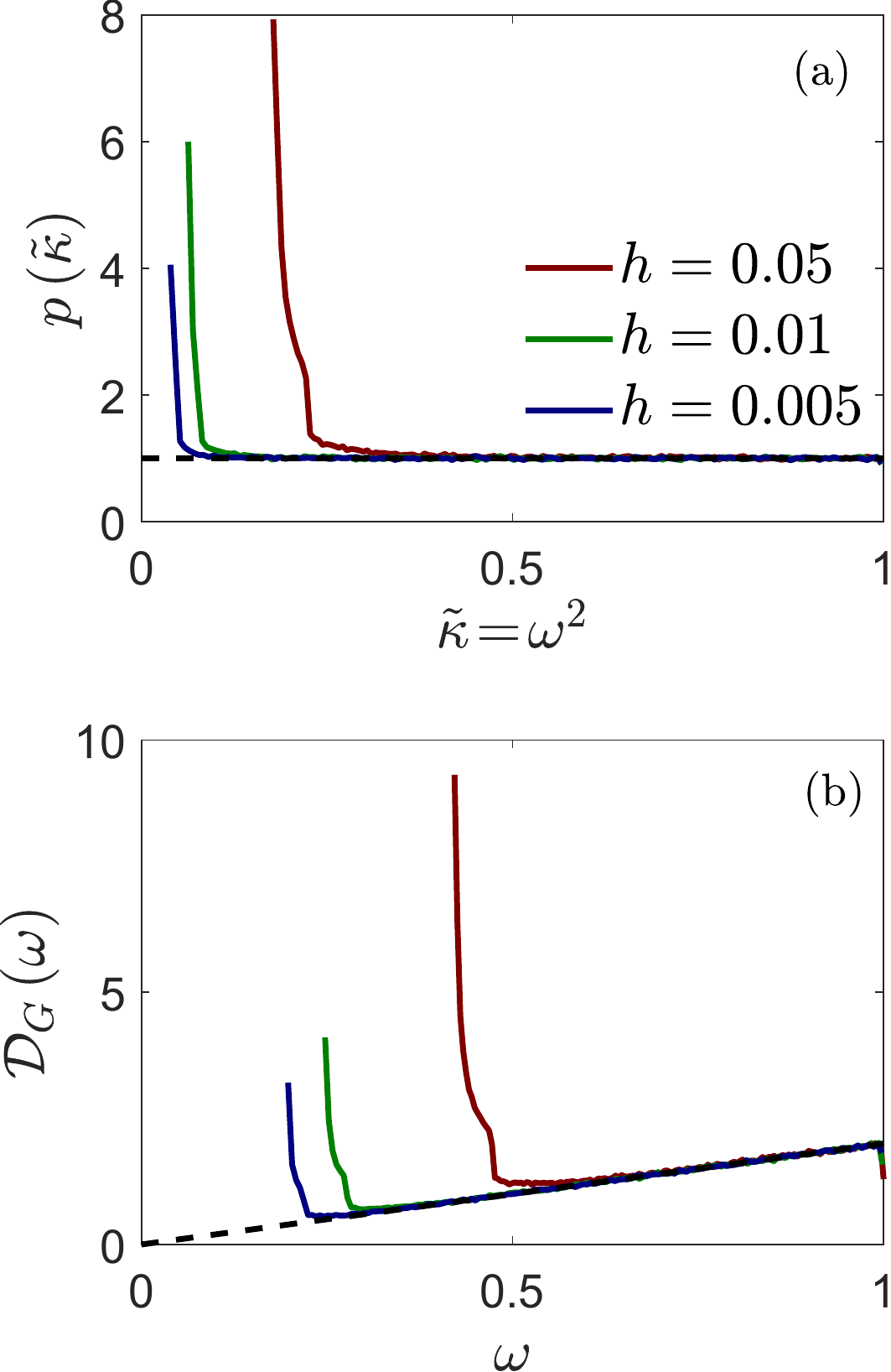}
\caption{(a) The transformation of the initial liquid-like distribution $p(\kappa)\!=\!\kappa_0^{-1}$ (for $0\!\le\!\kappa\!\le\!\kappa_0$, with $\kappa_0\!=\!1$) into $p(\tilde\kappa)$ upon the introduction of internal stresses represented by $h\!>\!0$, in the absence of interactions between oscillators, $J\!=\!0$ (i.e.~when the single-oscillator Hamiltonian reads $\kappa x^2/2 + x^4/24-hx$). $p(\tilde\kappa)$ is plotted for 3 values of $h$ (indicated in the legend), revealing a gap that increases as $\sim\!h^{2/3}$ and leads to the accumulation of $\sim\!Nh^{2/3}$ blue-shifted oscillators in a narrow stiffness range near $\tilde\kappa\!\sim\!h^{2/3}$. Note that $\tilde\kappa\!=\!\omega^2$ (as indicated in the $x$-axis label) and that the corresponding distribution for $\omega$ is shown in panel (b) for completeness. (b) The same as panel (a), but for the frequency $\omega$. That is, ${\cal D}_{\rm G}(\omega)$ for the non-interacting case of $J\!=\!0$ and $h\!>\!0$ is shown.}
\label{fig:figS3}
\end{figure}

In the manuscript, the main scaling predictions for $\bar{e}(\omega)$ --- i.e.~$\bar{e}(\omega_{\rm p})\!\sim\!h^{1/3} J$ and $\bar{e}_{\rm plateau}\!\sim\!J$ --- are derived using a perturbative approach in which the $J\=0$ and $h\!>\!0$ case is considered first, and then weak interactions $J\!>\!0$ are considered. It is stated therein that for $J\=0$ and $h\!>\!0$ (described by the non-interacting single-oscillator Hamiltonian $\kappa x^2/2 + x^4/24-hx$), the liquid-like vibrations below $\kappa\!\sim\!h^{2/3}$ are blue-shifted and accumulate in a narrow stiffness range near $\kappa\!\sim\!h^{2/3}$, while those above $\kappa\!\sim\!h^{2/3}$ are weakly affected. This is explicitly demonstrated in Fig.~\ref{fig:figS3}.
%%%%%%%%%%%%%%%%%%%%%%%%%%%%%%%%%%%%%%%%%%%%%%%%%%%%
%%%%%%%%%%%%%%%%%%%%%%%%%%%%%%%%%%%%%%%%%%%%%%%%%%%%

\section{Additional information and supporting results}

\subsection{Thermal history variability of various\\ physical quantities}

The experimental results summarized in Table~\ref{tab:exp} demonstrate the annealing variability of various basic physical quantities. It is observed that Debye's prefactor $A_{\rm D}$ decreases with annealing, which mostly reflects the stiffening of the elastic moduli. Indeed, the shear modulus $\mu$ increases with annealing (by $22\%$ between the most annealed and as quenched samples). As discussed extensively in this work, both $\nu_{\rm p}$ and ${\cal D}_{\rm G}(\nu_{\rm p})$ mildly increase with annealing. Finally, samples undergo densification (i.e.~increase of the mass density $\rho$) with annealing. Note in this context, that the experiments are done under NPT conditions (constant pressure) such that the thermal history affects the volume.

It would be interesting to consider also the co-variation of various physical quantities. The frequency scale $\nu_{\rm p}$ that is associated with the peak of ${\cal D}_{\rm G}(\nu)$ defines a stiffness scale $\sim\!\nu_{\rm p}^2$. The latter characterizes the stiffness of quasi-localized vibrations in the boson peak region, i.e.~it is a mesoscopic elastic response coefficient. According to Table~\ref{tab:exp}, $\nu_{\rm p}^2$ increases (stiffens) by $45\%$ (obtained from $(36.16/30.02)^2$) between the most annealed and as quenched samples. It would be natural and interesting to compare this observation to the annealing variability of the macroscopic elastic coefficient, i.e.~the shear modulus $\mu$. Previous work indicated that the degree of stiffening with annealing of $\mu$ is smaller compared to that of the mesoscopic elastic coefficient associated with quasi-localized vibrations~\cite{cge_paper,cge2_jcp2020,sticky_spheres1_karina_pre2021}. Indeed, the  annealing variability of $\mu$ in the experimental data summarized in Table~\ref{tab:exp} is $22\%$, a factor of 2 smaller than that of $\nu_{\rm p}^2$, as expected. These experimental observations are also consistent with the ideas and analysis of~\cite{wyart_vibrational_entropy}.

%%%%%%%%%%%%%%%%%%%%%%%%%%%%%%%%%%%%%%%
\begin{table*}[ht!]
    \centering
    \begin{tabular}{|c||c|c|c|c|c|}
    \hline
         Thermal history &
          $A_{\rm D}$  &
          $\omega_{\rm p}$  &
          $\mathcal{D}_{\rm G}(\omega_{\rm p})$ &
          $p$  & $\mu$ \\
        \hline    \hline
         $\dot{T}\!=\!\infty$  & $6.9\cdot10^{-4}$ (1) & 4.15 (1) &  0.0075 (1) &   18.88 (1) & 12.75 (1)  \\
         $\dot{T}\!=\!10^{-3}$ & $5.8\cdot10^{-4}$ (0.85) & 4.89 (1.18) &  0.009 (1.2) &  18.67 (0.99)  & 14.25 (1.12) \\
    \hline
    \end{tabular}
    \caption{The symbols of the different computer glass samples and the physical quantities that characterize them. The thermal history is characterized by the quench-rate used, see the `Thermal history' column. $\mu$ is directly computed, and $A_{\rm D}\!=\!3 / \omega_D^3$. $\omega_{\rm p}$ and ${\cal D}_{\rm G}(\omega_{\rm p})$ are extracted from Fig.~1d in the manuscript. Since the simulations are performed at a fixed volume, we report the hydrostatic pressure $p$ of the glasses (the counterpart of the density $\rho$ in Table~\ref{tab:exp}). The variation of each quantity, relative to the  $\dot{T}\!=\!\infty$ samples is reported in angular brackets in each column.}
    \label{tab:comp}
\end{table*}
%%%%%%%%%%%%%%%%%%%%%%%%%%%%%%%%%%%%%%%

The thermal history variability of ${\cal D}_{\rm G}(\omega)$ in the computer glass data, presented in Fig.~1d-e in the manuscript, bears close resemblance to the corresponding experimental data, presented in Fig.~1b-c in the manuscript. In Table~\ref{tab:comp}, we provide the actual values of the peak's location $\omega_{\rm p}$ and its magnitude ${\cal D}_{\rm G}(\omega_{\rm p})$, along with Debye's prefactor $A_{\rm D}\!=\!3/\omega^3_{\rm D}$, the hydrostatic pressure $p$, and the shear modulus $\mu$. $A_{\rm D}$ is obtained using Debye's frequency $\omega_{\rm D}$, computed using $\omega^3_{\rm D}\!=\!\frac{18 \pi^2 \rho}{2 v_{\rm t}^{-3} + v_{\rm l}^{-3}}$. Here, $v_{\rm t}\=\sqrt{\mu/\rho}$ is the transverse (shear) wave-speed and $v_{\rm l}\=\sqrt{(K\! +\! \tfrac{4}{3}\mu) / \rho}$ is the longitudinal (dilatational) wave-speed, where $K$ is the bulk modulus. Note that the simulations are done under NVT conditions (constant volume) such that the thermal history affects the pressure $p$, which we reported instead of the fixed mass density $\rho$.

The relative thermal history variability of the various physical quantities in Table~\ref{tab:comp} is reported in angular brackets in each column. These values make it easy to compare the relative thermal variability of various quantities in computer glasses presented in Table~\ref{tab:comp} to their experimental counterparts in Table~\ref{tab:exp}. The comparison reveals that despite the differences in composition and underlying interaction potential, and despite the different thermal history protocols (variable annealing in the experiments and variable quench rate in the computer simulations), the similarities in the nonphononic VDoS observed in Fig.~1 in the manuscript are semi-quantitatively echoed in the two Tables.

\subsection{Thermal history variability of dimensionless quantifiers of mechanical disorder}

Computer glass simulations, in view of their atomistic resolution, provide access to physical quantities that are currently not accessible experimentally. In recent years, several dimensionless quantifiers of mechanical disorder in glasses have been developed and substantiated~\cite{phonon_widths2,karina_chi_paper_2023,lutsko,lemaitre2004}. These dimensionless quantifiers allow to put on equal footing different glasses and compare their degree of mechanical disorder. It would therefore be useful to report the values of these dimensionless quantifiers for the computer glasses that have been compared to experiments in the previous subsection.

To that aim, we prepared ensembles of a few thousand glass samples of a few thousands of particles each. These are needed for statistical convergence and were used to compute three dimensionless quantifiers of mechanical disorder, as reported next. The first quantifier --- studied and discussed in detail in~\cite{phonon_widths2,karina_chi_paper_2023} --- captures the sample-to-sample fluctuations of the macroscopic shear modulus $\mu$; it is defined as
\begin{equation}
\chi \equiv \sqrt{N}\frac{\mbox{STD}(\mu)}{\mbox{MEAN}(\mu)}\,,
\end{equation}
where the standard deviation and mean appearing above refer to \emph{ensemble} averages. We find $\chi(\dot{T}\!=\!10^{-3})\!\approx\!3.9$ and $\chi(\dot{T}\!=\!\infty)\!\approx\!2.6$, see~\cite{karina_chi_paper_2023} for a comparison of these values with a wide variety of disordered solids.

The second quantifier is the ratio of the nonaffine contribution to the shear modulus that emerges due to glasses' structural disorder/frustration~\cite{lutsko,lemaitre2004}, and the total shear modulus, namely $\mu_{\rm na}/\mu$. Detailed definitions of $\mu_{\rm na}$ and $\mu$ can be found e.g.~in~\cite{lutsko,lemaitre2004}, and some representative values of generic computer glasses can be found in~\cite{disordered_crystals_prl_2022}. We find $\mu_{\rm na}/\mu\!\approx\!1.05$ for our $\dot{T}\!=\!10^{-3}$ ensembles, and $\mu_{\rm na}/\mu\!\approx\!1.32$ for our hyperquenched $\dot{T}\!=\!\infty$ samples.

The third and last dimensionless quantifier of mechanical disorder we report for our computer glasses is the product $A_{\rm g}\,\omega_\star^5$, where $A_{\rm g}$ is the prefactor of the $\sim\!\omega^4$ scaling regime of the nonphononic VDoS, and $\omega_\star\!\equiv\!v_{\rm t}/a_0$ is a characteristic elastic frequency. Here, $a_0\!\equiv\!(V/N)^{1/3}$ is a characteristic interparticle distance. Typical values for the product $A_{\rm g}\,\omega_\star^5$ can be found in e.g.~\cite{pinching_pnas,disordered_crystals_prl_2022,sticky_spheres1_karina_pre2021,karina_chi_paper_2023}. Here, we find $A_{\rm g}\,\omega_\star^5\!\approx\!0.50$ for our $\dot{T}\!=\!10^{-3}$ ensembles, and $A_{\rm g}\,\omega_\star^5\!\approx\!1.35$ for our hyperquenched $\dot{T}\!=\!\infty$ samples.

%\bibliography{glass_refs_2023}
%\bibliographystyle{unsrt}

\end{document}